# INVESTIGATION OF THERMAL ACOUSTIC EFFECTS ON SRF CAVITIES WITHIN CM1 AT FERMILAB*

M.W. McGee[†], E. Harms, A. Klebaner, J. Leibfritz, A. Martinez, Y. Pischalnikov, W. Schappert,
Fermi National Accelerator Laboratory, Batavia, IL 60510, USA

## Abstract

Radio Frequency (RF) power studies are in progress following the cryogenic commissioning of Cryomodule #1 (CM1) at Fermilab's Superconducting Radio Frequency (SRF) Accelerator Test Facility. These studies are complemented by the characterization of thermal acoustic effects on cavity microphonics manifested by apparent noisy boiling of helium involving vapor bubble and liquid vibration. The thermal acoustic measurements also consider pressure and temperature spikes which drive the phenomenon at low and high frequencies.

## INTRODUCTION

A 1.3 GHz, type III+ European X-Ray Laser Project (XFEL) cryomodule consists of eight dressed 9-cell niobium superconducting radio frequency (RF) cavities. The cold mass hangs from three column support posts constructed from G-10 fiberglass composite, which are attached to the top of the vacuum vessel. The 312-mm diameter helium gas return pipe (HeGRP), supported by the three columns, acts as the coldmass spine, supporting the cavity string, quadrupole and ancillaries. A 2-phase pipe (manifold) delivers helium flow to each cavity helium vessel [1-4].

Brackets with blocks on two sides provide a connection between each cavity and the HeGRP. Two aluminum heat shields (80 K and 5 K) hang from the same two column supports. The coldmass consists of all components found within the 80 K shield shown in Fig. 1. Relative longitudinal and transverse alignment (or position) of the cavity string and quadrupole is held by an Invar rod (a material with very low thermal expansion).

The helium vessel has a large chimney nozzle for CW heat removal with an upstream US located cool-down / warm-up supply port. Each cryogenic manifold is constructed of stainless steel 316L. Pressure drops within and through the module have been portioned in combination with the helium distribution system. These pipes were sized for the worst case among steady-state, peak flow rates, upset, cool-down, warm-up, and venting conditions [5].

## HELIUM FILM BOILING

Of the two types, noiseless and noisy film boiling, superfluid helium is the most interesting. Boiling in the noiseless regime has a smooth and wavy interface boundary, somewhat constant in size and time and vapor does not form in the film layer. High frequency noise (similar to the acoustic noise occurring when conventional subcooled liquids boil) is generated by noisy superfluid film. The volume of vapor bubbles constantly grow and then collapse producing the characteristic noise. Between those vapor bubbles, sections of smooth cylindrical interface exist as in the noiseless film boiling regime. Many research sources suggest that the heat transfer in the film boiling regime is less than in the noiseless regime.

A layer of 'normal' thermal conductivity with a temperature near saturation corresponding to the vapor film pressure causes small vapor bubbles. In Figure 2, solid circles represent vapor bubble oscillations; the unfilled circles the liquid oscillation and the solid lines gives the theoretical prediction of nth mode of liquid column oscillation. This involved superfluid helium at 2.15 K [6].

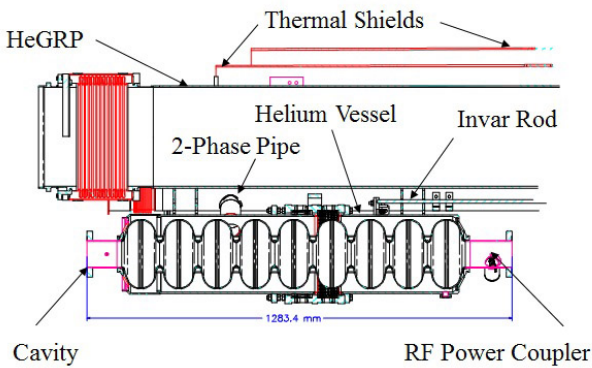

Figure 1: Elevation view of CM1, cavity #1.

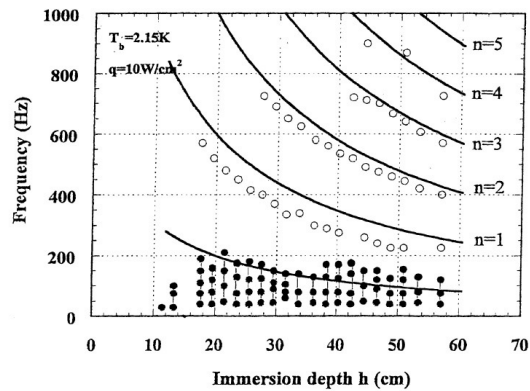

Figure 2: Frequency versus pressure oscillations [6].

## MEASUREMENT

A seismic station was established centered with respect to CM1 within the cave at New Muon Lab (NML) and set back between staggered shielding blocks. This station consists of a vertical (y), transverse (x) and longitudinal HS-1 geophone (NML REF) and a vertical Sercel Mark L-4c seismometer (NML REF Ser). The seismic station established helped define external input such as ground motion.

### Instrumentation

CM1 was instrumented with Oyo Geospace GS-14-L9 geophones [7] mounted on a supporting bracket above cavity #3 (Cav3), #5 (Cav5) and #7 (Cav7), one vertical and one horizontal. These geophones were enclosed within a cylindrical magnetic shield made of heat-treated 1018 low-carbon steel to protect the cavities from stray magnetic fields. A vertical Oyo Geospace GS-11-D geophone [8] and horizontal geophone SM6-HB from Sensor b.V [9] were attached on the quad, upstream. Each device was connected to a National Instruments (NI) NI-9233 4-channel, 24-bit ADC module sampled at 1,500 K/sec, and the data was collected using a dedicated PC desktop.

## LIQUID HELIUM OPERATION

The cooldown began in mid November of 2010 as turbulent gas flow of initial circulation of cold nitrogen and helium gas within the cryomodule CM1 piping circuit. The first measure of liquid helium at 4.2 K formed two days later within the cavity string of cryomodule CM1 on November 19[th]. Erratic helium gas flow with random pressure spikes caused the cavities, quad, and HeGRP to resonate. This behavior quickly became quiescent as liquid helium began to form and fill each cavity [10].

### Superfluid Operation

Superfluid helium within the cryomodule CM1 cavity string was achieved during the initial operation. Vertical and horizontal acceleration and motion measurements taken during operation at 2 K (23.4 Torr) reflect the stability of cryomodule CM1 without RF power. Once RF power with 5 Hz pulse rate was introduced, initial integrated displacement (rms) measurements indicated 20 μ motion for the quad and average cavity motion was 2 μ. Accelerations during helium boiling within the cavity is consider in terms of motion in Figure 3. Cavity motion increased to 10 μ and quad motion reached 100 μ (rms) during operational boiling.

A magnitude change of the downstream (DS) quad vertical motion was significant. This response was driven by the boiling within each cavity helium vessel alone. This motion is much higher than the vertical response of 22 μ (rms) during initial cooldown.

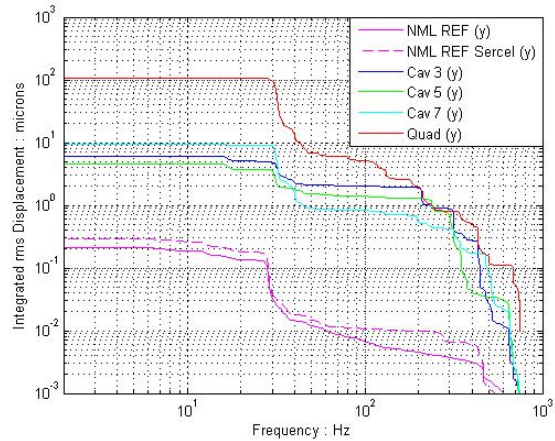

Figure 3: Vertical integrated displacement (rms).

## SUPERFLUID BOILING

Beam and cryogenic system operation combine to produce conditions leading to boiling superfluid helium surrounding each cavity. Vertical accelerations increasing to 0.35 g were observed during helium boiling, ringing the HeGRP as shown in Figure 4. Note that the DS dummy quad also experienced the acceleration. Normal operation liquid helium flow rate measurements were conducted from 0 to 6 g/s, where the cavity displacement (rms) was found to be constant [10]. Excessive RF power heating caused the boiling, given the same operating pressure, temperature and flow.

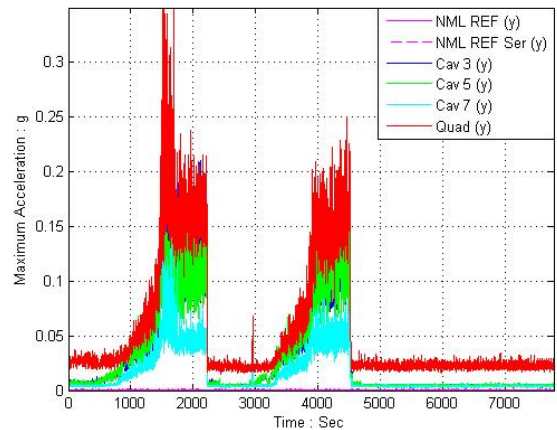

Figure 4: Vertical acceleration during cavity boiling.

The red colored (active) regions within the spectrogram (shown in Figure 5) represents the low-frequency vapor bubble oscillations found at or below 210 Hz. These low-frequencies cover much of the spectrum from 16 Hz and increasing above 210 Hz. An overlapping transition is observed near 210 Hz from vapor bubble to liquid oscillations. The periods of superfluid boiling are apparent from ~ 1,000 to 2,200 seconds and from ~ 3,000 to 4,500 seconds.

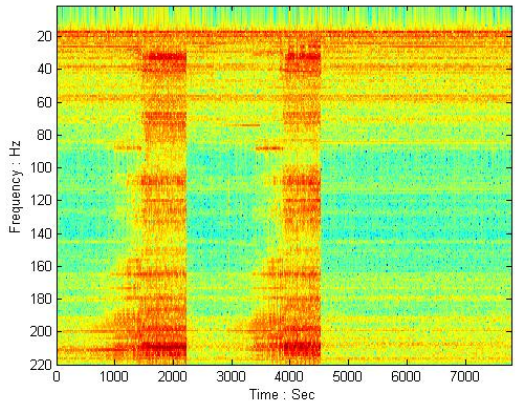

Figure 5: Vertical spectrogram showing vapor bubbles.

Higher frequency superfluid helium liquid oscillations occurred at different multiples (n = 1, n = 2, etc.) of 210 Hz, depicted as black lines in Figure 6. Under the conditions of 2.15 K and heat rate of 10 Watts/cm$^2$ (see Figure 2), the immersion depth of superfluid boiling is very shallow within each cavity helium vessel ~ 15 cm.

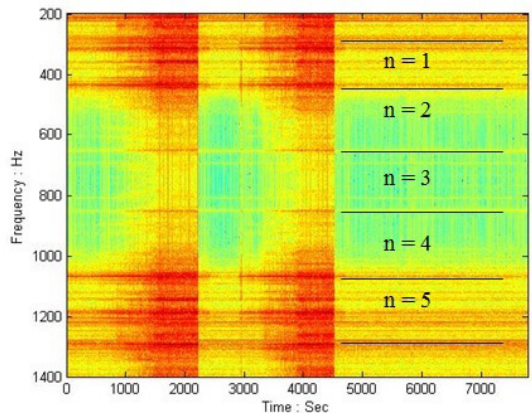

Figure 6: Vertical spectrogram showing liquid oscillations.

Also, given the pressure-temperature liquid boiling stability curve shown in Figure 7, the superfluid boiling is unstable and approaches the vapor saturation line.

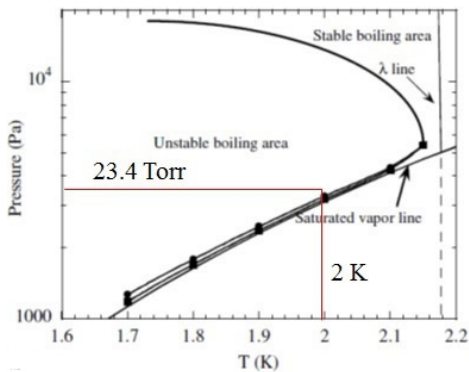

Figure 7: Stability Pressure – Temperature curve [11].

## FUTURE WORK

Continued studies have been considered regarding the current Cryomodule #2 (CM2) operation within the cave at NML. Fast Fourier Transform (FFT) analysis coupled to our ACNET system has provided a developing diagnostic tool. Further tests will help to refine these defined parameters of peak FFT response.

## ACKNOWLEDGEMENTS

We wish to thank Jirawat (Mack) Amorn-Vichet, Ryan Heath, Stewart Mitchell, Tony Parker, Dave Slimmer and Nino Strothman (Computer, Networking and Labview Application Support Personnel). Also, thanks to Brian DeGraff and Greg Johnson for cryogenic support. Special thanks to Kermit Carlson and Wayne Johnson for their technical support.